\pdfoutput=1
\documentclass[12pt]{article}
\usepackage{jheppub}

\usepackage[utf8]{inputenc}

\usepackage{amsmath}
\usepackage{amssymb}
\usepackage{graphicx,color}
\usepackage{amsthm}
\usepackage{mathrsfs}

\usepackage{tikz}
\usetikzlibrary{matrix,arrows}

\usepackage{ifpdf}

\usepackage{subfigure}

\usepackage[boxsize=0.5em,aligntableaux=center]{ytableau}

\usepackage{tikz}
\usetikzlibrary{shapes,shadows,calc}
\usepgflibrary{arrows}
\usepackage{rotating}
\usepackage{wrapfig}

\tikzset{
  sshadow/.style={opacity=.25, shadow xshift=0.05, shadow yshift=-0.06},
}

\def\kbbox[#1,#2,#3,#4,#5]#6{
        \draw[dashed] node[draw,color=gray!50,minimum
        height=#1,minimum width=#2] (#4) at #5 {}; 
        \node[anchor=#3,inner sep=2pt] at (#4.#3)  {#6};
}
\def\kbboxred[#1,#2,#3,#4,#5]#6{
        \draw[] node[draw,color=red,minimum
        height=#1,minimum width=#2] (#4) at #5 {}; 
        \node[anchor=#3,inner sep=2pt] at (#4.#3)  {#6};
}

\newcommand{\cH}{\mathcal{H}}

\newcommand{\cM}{\mathcal{M}}
\newcommand{\cN}{\mathcal{N}}

\newcommand{\cS}{\mathcal{S}}
\newcommand{\cT}{\mathcal{T}}

\newcommand{\cG}{\mathcal{G}}

\newcommand{\bbZ}{\mathbb{Z}}
\newcommand{\bbR}{\mathbb{R}}

\DeclareMathOperator{\re}{Re}

\def\a{{\alpha}}
\def\b{{\beta}}
\def\eps{{\epsilon}}
\def\th{{\theta}}
\def\Lam{{\Lambda}}
\def\lam{{\lambda}}

\def\g{{\gamma}}

\def\p{{\partial}}

\def\bs#1\es{\begin{split}#1\end{split}}
\def\ba#1\ea{\begin{align}#1\end{align}}
\def\baed#1\eaed{\begin{aligned}#1\end{aligned}}
\def\bged#1\eged{\begin{gathered}#1\end{gathered}}

\def\nn{\nonumber}

\def\a{\alpha}
\def\b{\beta}

\def\g{\gamma}

\let\foo\bar 
\renewcommand{\bar}[1]{ {\foo{  #1} }{} }

\newlength{\dhatheight}

\def\Label#1{\label{#1}%
  \smash{\hbox to0pt{\raise1ex\hbox{\tiny[#1]}\hss}}}
\def\noLabels{\let\Label=\label}
\def\nobbibitem{\let\bbibitem=\bibitem}
 \def\noBibitem{\let\Bibitem=\bibitem}

\newcommand{\be}{\begin{equation}}
\newcommand{\ee}{\end{equation}}
\newcommand{\beq}{\begin{equation}}
\newcommand{\eeq}{\end{equation}}
\newcommand{\bea}{\begin{eqnarray}}
\newcommand{\eea}{\end{eqnarray}}

\newcommand\varpm{\mathbin{\vcenter{\hbox{%
  \oalign{\hfil$\scriptstyle+$\hfil\cr
          \noalign{\kern-.3ex}
          $\scriptscriptstyle({-})$\cr}%
}}}}
\newcommand\varmp{\mathbin{\vcenter{\hbox{%
  \oalign{$\scriptstyle({+})$\cr
          \noalign{\kern-.3ex}
          \hfil$\scriptscriptstyle-$\hfil\cr}%
}}}}

\title{\centering \Large On dualities for non-Abelian gauge theories with continuous center}

\author[]{Thomas W.~Grimm}
\author[]{and Diego Regalado}

\affiliation[]{Max Planck Institute for Physics,\\
F\"ohringer Ring 6, 80805 Munich, Germany\\[.1cm]
\text{\normalfont and}}

\affiliation[]{
Institute for Theoretical Physics and \\
Center for Extreme Matter and Emergent Phenomena,\\
Utrecht University, Leuvenlaan 4, 3584 CE Utrecht, The Netherlands}

\emailAdd{grimm@mpp.mpg.de}
\emailAdd{regalado@mpp.mpg.de}

\abstract{Formulating gauge theories for gauge groups admitting a continuous center
 can require to include charged scalars to define 
a gauge-coupling function. We show that the gauge-fields in the center can be 
dualized into form-fields of dimension-dependent degree. The resulting theory 
admits a smaller gauge group that factorizes out the center, but contains a Chern-Simons type term coupling 
the scalars to the form-fields. 
As an explicit example we consider the gauge group being the Heisenberg group
and show that the dual action only admits an Abelian gauge symmetry. 
We comment on the vacuum symmetries in these settings, their 
supersymmetrization, and 
point out their importance in string theory.}

\begin{document}
\setlength{\parskip}{5pt}

\makeatletter
\renewcommand{\@fpheader}{%\old@fpheader
\hfill MPP-2015-164}
\makeatother

\maketitle
\newpage

\section{Introduction}

A classical duality generalizing four-dimensional electro-magnetic duality is 
provided by the fact that in $d$-dimensions a massless Abelian vector admits a 
dual description in terms of a massless $(d-3)$-form. In order that such a 
dual theory exists it is crucial that the vector field only appears with additional 
derivatives in the action. Clearly, this is generally not true for non-Abelian gauge 
theories, since the bare gauge-fields can appear in the definition of the field strength.
For certain classes of non-Abelian groups, however, a subset of the gauge-fields 
only occurs together with derivatives. Such groups are exactly 
the ones with continuous center, since by definition all gauge fields parameterizing the 
center commute with all other group elements. In this work we focus on such groups 
and study the associated gauge theory. We will also show that in this 
cases the duality to a description with $(d-3)$-forms can be performed. These 
dual descriptions are often crucial when studying effective actions of string theory. 

Already the Lagrangian formulation of gauge theories with the gauge group admitting 
a continuous center is more involved than for standard Yang-Mills theories. 
This can be traced back to the fact that their Killing form has non-maximal rank and a
simple kinetic term for the gauge-fields cannot be defined using the Killing form alone. 
To nevertheless define a kinetic term one is required to 
include extra scalar degrees of freedom that transform appropriately 
under the gauge group. One can then naturally introduce a 
field-dependent gauge-coupling function that is positive definite and
non-vanishing along certain parts of the scalar field space. 
As expected this kinetic term will only depend on the derivative of gauge-fields along 
the center of the group, while the bare gauge fields from the center do not appear. 
This suggests that there indeed exists a dual description involving $(d-3)$-forms.
As we will discuss in detail in this work this duality can be more involved due to 
required presence of the charged scalars. Treating the gauge fields in the center similarly to 
massive fields we find that nevertheless a dual description can be established. Remarkably, 
this dual formulation will only admit a smaller gauge group supplemented by Abelian local 
symmetries for the form fields. 

This work can be motivated from various directions. Firstly, one 
realizes that the Heisenberg group is a key example for a group with 
continuous center. Generalizations of this group appear often 
as symmetry groups of string theory moduli spaces. Therefore one might 
wonder how these symmetry groups can be consistently gauged and 
how such gaugings can be detected in the effective action. Effective 
actions with gauged Heisenberg groups arise, for example, in 
compactifications of Type II string theory or M-theory on five- or six-dimensional 
manifolds with $SU(2)$ structure \cite{Cassani:2010uw,Liu:2010sa,Gauntlett:2010vu,Danckaert:2011ju,KashaniPoor:2013en,Grimm:2014aha}, 
or M-theory on manifolds with $SU(4)$ structure \cite{Grimm:2015ona}. 
Other reductions with non-Abelian gaugings of this type are reviewed, for example, in \cite{Samtleben:2008pe}.
In such 
reductions the effective action might not immediately be in the correct 
frame to infer the full non-Abelian gauge group and several dualization 
steps have to be performed. Our work explains in detail what happens to 
the gauge groups in such dualizations and highlights the occurring curiosities. For 
example, while in the vector formulation with Heisenberg group one is dealing with a
non-Abelian gauge theory, the dual theory with $(d-3)$-forms might only admit an 
Abelian gauge symmetry. 

A second motivation is provided when studying the vacuum configurations
in gauge theories with continuous center. In fact, due to the required presence 
of gauged scalars to define the theory, their vacua can have interesting discrete 
Abelian and non-Abelian gauge symmetries.
Recall that discrete Abelian gauge symmetries such as $\bbZ_p$ can be understood as arising 
by starting with a $U(1)$ gauge group and gauging a Higgs scalar in a non-linear fashion. 
In a dual description one can replace the $U(1)$-gauge field by a 
$(d-3)$-form and the Higgs scalar by a $(d-2)$-form. This dual theory admits an
emergent $U(1)$ gauge symmetry (see \cite{Banks:2010zn} for an in-depth discussion on such discrete Abelian 
symmetries). The non-Abelian generalization of this discussion arises naturally for 
gauge groups with continuous center. For example, starting with 
the continuous Heisenberg group the required coupled Higgs scalars can 
break it to the discrete Heisenberg group $H_{\bbZ}$ in the vacuum. 
Replacing the gauge fields of the center with dual $(d-3)$-forms and the coupling
Higgs scalars by $(d-2)$-forms the resulting theory has an emergent symmetry 
group just as in the Abelian case. In contrast to the Abelian case, the emergent 
symmetry group generally differs from the original non-Abelian group. 

The paper is organized as follows. In section \ref{dual_Heisenberg} we discuss the gauge theory 
for the three-dimensional Heisenberg group. We argue that additional scalars are required 
to define a kinetic term, which at the same time allow for the vacuum symmetry group to 
be a discrete non-Abelian subgroup of $H$. The dualization of the vector spanning the center of 
$H$ is performed in detail. The generalization of this construction to other groups with continuous center 
can be found in section \ref{sec:generaldis}. Again we first construct the non-Abelian gauge theory 
and then perform the dualization of the vectors parameterizing the center of the gauge group. 
We conclude in section \ref{sec:conclusions} by commenting on the significance of the discussed gauge 
groups in string theory and provide a brief discussion on the supersymmetrization of the constructed 
actions.

%%%%%%%%%%%%%%%%%%%%%%%%%%%%%%%%%%%%%%%%%%%%%%%%
\section{Dual Heisenberg actions and discrete symmetries} \label{dual_Heisenberg}
%%%%%%%%%%%%%%%%%%%%%%%%%%%%%%%%%%%%%%%%%%%%%%%%
In this section we introduce gauge theories that admit the three-dimensional 
Heisenberg group $H$ as gauge group. This first simple example will highlight many 
key features of how to formulate a gauge theory for groups with continuous center. 
The Heisenberg gauge theory itself is introduced in subsection \ref{Heisenberg_gaugetheory}, 
where we will also comment on the discrete remnants of this group when considering 
vacuum configurations.
In subsection \ref{sec_dualHeis} we perform the dualization of the vector parameterizing the center of $H$ 
into a $(d-3)$-form. We discuss the symmetries of the dual action and show that it admits 
only Abelian gauge symmetries.

\subsection{Heisenberg gauge theory and non-Abelian 
discrete symmetries} \label{Heisenberg_gaugetheory}

We first construct the action describing a gauge theory 
with the Heisenberg group $H$ as gauge group in $d$ space-time 
dimensions. We consider the
simplest case in which $H$ is three-dimensional. It 
takes the form
\beq
    H =  \mathbb{R} \ltimes (\cH_1 \times \cH_2)\, ,
\eeq
where $\cH_i$ is either $\bbR$ or $U(1)$.
$H$ is generated by three algebra 
elements $t_A$ with $A=1,2,3$ satisfying
\beq
[t_1,t_2] = - Mk \, t_3\ ,
\eeq 
with all other commutators vanishing. 
The only non-trivial structure constants are $f_{ab}^3 = - Mk \epsilon_{ab}$, 
where $\epsilon_{ab}$ is the two-dimensional Levi-Civita symbol and $a,b=1,2$. 

Next we introduce the gauge fields $A^A$ for the generators $t_A$
that transform under the gauge symmetry as
\bea \label{Heisenbergaction}
\delta A^a &=& d\lam^a \, , \\
\delta A^3&=& d\lam^3+Mk\epsilon_{ab}A^a\lam^b-\frac{1}{2}Mk\epsilon_{ab}\lam^a d\lam^b \, . \nn
\eea
Note that this is the transformation law for actual finite group actions.
At infinitesimal level only the first two terms in $\delta A^3$ are considered.
The field strengths of $A^A$ are denoted by $F^A = dA^A - \frac12 f_{BC}^A A^B \wedge A^C$.
One readily checks that gauge invariance of 
their kinetic term
\beq \label{Skin1}
    S^{(d)}_{\rm kin} = -  \int   Q_{AB} F^A \wedge * F^B 
\eeq
implies that $Q_{AB}$ has to transform non-trivially under $H$.
More precisely, one has
\beq \label{Q_trans}
   Q_{AB} \ \rightarrow \  Q_{CD} (D^{-1})_A^C (D^{-1})_B^D\ ,
\eeq
where $D_A^B$ is the adjoint of the group. For the Heisenberg group \eqref{Heisenbergaction}
one readily computes 
\beq
   (D_A^B) = \left(\begin{array}{ccc}1& 0 & 0 \\ 
                                             0 & 1 & 0 \\ 
                                             M k \lambda^2 & - M k \lambda^1 & 1 \end{array}  \right)\, ,
\eeq
which is independent of $\lambda^3$. 
Note that the Heisenberg group has no positive-definite 
Killing form. To nevertheless find a positive-definite $Q_{AB}$ 
one can introduce scalar fields 
charged under $H$ and make $Q_{AB}$ field-dependent.

There are two ways that this can be achieved which we discuss in the following.
The minimal approach is to introduce two periodic scalars $b^a$
that transform under \eqref{Heisenbergaction} as
\beq \label{transbi}
  \delta b^a= k\lam^a\, , \qquad  b^a \cong b^a + 1\, ,
\eeq
and admit gauge invariant derivatives 
\beq
 D b^a = d b^a-kA^a\, .
\eeq
The periodicity ensures that we 
find interesting vacuum configurations and 
allows us to connect with concrete string theory 
realizations. 
A gauge-coupling function $Q_{AB}$ rendering
 \eqref{Skin1} invariant then takes the form 
 \beq
  (Q_{AB}) = \left(\begin{array}{cc}  \cM_{ab}+ (M)^2\epsilon_{ac}\epsilon_{b d} b^c b^d & M\epsilon_{c a} b^c \\ 
  M\epsilon_{c b} b^c & \mathcal{M} \end{array} \right)\ ,
 \eeq
 where $\cM_{ab}$ is positive definite and $\cM$ is a positive constant.
With these definitions at hand  a gauge-invariant 
action is given by
\beq \label{Skin2}
    S^{(d)}_{\rm min} = -  \int   Q_{AB} F^A \wedge * F^B
                      + \cN_{ab} D b^a \wedge *Db^b\, ,
\eeq 
where $\cN_{ab}$ is positive definite and independent of $b^a$. 

The space of inequivalent vacua of \eqref{Skin2} is obtained for $A^A=0$ and constant scalars $\langle b^a\rangle$.
Constant gauge transformations preserving the $\langle b^a\rangle$ correspond to a preserved vacuum symmetry.
The periodicity of the $b^a$ in \eqref{transbi} yields to a breaking of the continuous Heisenberg symmetry $H$ to  
\beq
     G_{1} = \mathbb{Z}_k \ltimes (\cH_1 \times \mathbb{Z}_k )\ .
\eeq
Note that this is a non-compact group for $\cH_1=\bbR$ and as such not expected to arise 
in a theory of quantum gravity. In contrast $G_1$ is a compact group for $\cH_1=U(1)$.

One might wonder if \eqref{Skin2} can be extended such 
that the scalars break the symmetry to the discrete Heisenberg 
group $H_{\mathbb Z}$.
In order to do that on has to introduce a third periodic 
scalar $b^3$ that transforms under the group $H$ as
\beq \label{transb3}
 \delta b^3 = p\lam^3+\frac{M p}{2}\epsilon_{ab} b^a \lam^b \, ,\qquad b^3 \cong b^3+1\, ,  
\eeq  
with the transformations of the $b^i$ in \eqref{transbi} unchanged. 
The covariant derivatives $b^3$ is given by
\beq \label{Db3}
  D b^3  = db^3 -pA^3-\frac{Mp}{2}\eps_{ab}b^a A^b\,.   % +\frac{Mp}{2k}\epsilon_{ij}b^iDb^j
\eeq
The full gauge-invariant 
action now takes the form 
\beq \label{Skin3}
    S^{(d)}_{\rm kin} = -  \int  Q_{AB} F^A \wedge * F^B  
                      + \tilde \cT_{AB} Db^A \wedge *Db^B\, ,
\eeq 
where 
\beq
  (\tilde \cT_{AB}) = \left(\begin{array}{cc} \mathcal{N}_{ab}+\frac{(Mp)^2}{4 k^2 \cG }\epsilon_{a c} \epsilon_{b d} b^c b^d & \frac{M p}{2  k \cG}\epsilon_{c a}b^c \\ \frac{M p}{2   k \cG}\epsilon_{c b}b^c &  \frac{1}{\cG}\end{array} \right)\,.
 \eeq
Here $\cG$ is a positive constant.

The symmetry group of the vacua of \eqref{Skin3} is now the discrete Heisenberg group $H_{\bbZ}$,
\beq
   H_{\bbZ}  = \bbZ_k \ltimes (\bbZ_{p} \times \bbZ_{k})\ .
\eeq
Note 
that this group is compact. The appearance of $H_{\bbZ}$ in string theory is therefore in accordance with the 
`folk theorem' that forbids non-compact gauge symmetries in a theory of quantum gravity. 

To close this subsection we note that the states that can directly couple 
to the fields in \eqref{Skin3} are simply particles. They can be charged under 
(the discrete remnants) of the Heisenberg group $H$. However, we know from general arguments about discrete
gauge symmetries \cite{Lee} that there are also charged $(d-1)$-branes that couple minimally to the dual $(d-2)$-forms
of the scalars $b^a$.

\subsection{Dualizing the action} \label{sec_dualHeis}

In the next step we aim to study a dual frame describing the theories 
\eqref{Skin2} and \eqref{Skin3}. This will yield us to discover emergent 
new gauge symmetries and a gauge group $G_e$ that generally 
differs from all groups encountered so far. 

Our starting point is the observation that the vector $A^3$ only
appears with derivatives in the action \eqref{Skin2}. This implies 
that it should be possible to replace it by a dual degree of freedom
which is given by a $(d-3)$-form $B$. This dualization proceeds 
in the standard way. Firstly, we replace $dA^3$ with $\tilde F^3$
everywhere in \eqref{Skin3}. Then we add a Lagrange multiplier
term to arrive at the parent action
\beq \label{Spar1}
   S^{(d)}_{\rm par} = S^{(d)}_{\rm min}(dA^3\rightarrow \tilde F^3) + 2 \int d B \wedge \tilde F^3\ .
\eeq
The equations of motion for $B$ simply imply $d \tilde F^3=0$, such 
that locally one can write $\tilde F^3 = dA^3$. Inserted back into \eqref{Spar1}
we arrive back at the original action \eqref{Skin2}. However, we 
can also choose to eliminate $\tilde F_3$ and keep $B$. 
Before doing this, let us comment on the symmetries of \eqref{Spar1}. 
The fundamental fields in \eqref{Spar1} are $A^i$, 
$b^i$ and $B$, $\tilde F^3$. The symmetry transformations for 
$\tilde F^3$ are identical to those of $dA^3$ and 
inferred from \eqref{Heisenbergaction}. They capture the non-Abelian 
structure of $H$ and are a symmetry of \eqref{Spar1} up to 
a total derivative. Note that in contrast $B$ has 
only an Abelian symmetry. After eliminating $\tilde F^3$
from \eqref{Spar1} the the dual action reads
\bea \label{Sdual1}
   S^{(d)}_{\text{min-e}} &=&-  \int \cM_{ab} d A^a \wedge * dA^b  
                      + \cN_{ab} Db^a \wedge *Db^b  + \cM^{-1} \, dB \wedge * dB \\ 
                      &&   \qquad + \frac{M}{k} \epsilon_{ab}\, dB \wedge Db^a \wedge Db^b\, .\nn
\eea
Note that seemingly the non-Abelian structure of $H$ has disappeared, since all the 
covariant derivatives are Abelian and an Abelian symmetry of $B$
has emerged. Furthermore, \eqref{Sdual1} contains a particular new Chern-Simons 
type term. The seemingly higher-derivative part of this term is simply a total derivative 
and the chosen form allowed us to highlight its gauge invariance. As we will see 
in the remaining parts of this work, precisely terms of this type signal 
the presence of a non-Abelian gauge group in the dual description. 

Let us now turn to the discussion of the second action \eqref{Skin3}
that admits the discrete Heisenberg group $H_{\bbZ}$ as 
vacuum symmetry. In this case also $A^3$ is appearing without 
derivatives in the covariant derivative \eqref{Db3} 
of the scalar $b^3$. This implies that 
$A^3$ is massive and has to be dualized into a massive form field. 
In other words, a general dualization of $A^3$ into $B$ can now only be performed 
when dualizing $b^3$ at the same time into a $(d-2)$-form $V_{d-2}$.
The parent action for \eqref{Skin3} now takes the form 
\bea \label{Spar2}
S^{(d)}_{\rm par}&=& - \int \tilde \cT_{AB} Db^A \wedge*Db^B +  \cM_{ab}dA^a\wedge *dA^b  \\
     && \qquad \qquad +  {\mathcal M}^{-1}\, DB\wedge*DB +2 DB\wedge U \, , \nn 
\eea
where 
\bea
  DB&=&dB-pV  \\
  U&=&dA^3+\frac{Mk}{2}\epsilon_{ab} A^a \wedge A^b+M \epsilon_{ab} b^a dA^b\, . \nn 
\eea
It is interesting to notice that this parent action has the 
local symmetries given by \eqref{Heisenbergaction}, \eqref{transbi}, and \eqref{transb3}
supplemented by 
\beq
\delta V =d \Lambda \, , \qquad \delta B=p\Lambda\, ,
\eeq
where $\Lambda$ is a $(d-3)$-form. 
This corresponds to the group $H \times U(1)_{d-3}$.\footnote{The subscript in $U(1)_{d-3}$ is included to stress that the corresponding gauge parameter is a $(d-3)$-form.}
 Notice that the transformation rules given above are 
valid for finite gauge parameters $\lam$ and that $U$ is gauge-invariant. 
The parent action \eqref{Spar2} is build to precisely yield the action \eqref{Skin3} upon using 
the equations of motion of $V$ and $B$ to eliminate these fields from the theory. 

Alternatively we can use the equations of motion of $b^3$, $A^3$ and eliminate these 
degrees of freedom from the action. 
The result is the theory
\bea\label{Sdual2}
S^{(d)}_{\rm e}&=&- \int \cN_{ab} Db^a\wedge*Db^b+ \cM_{ab}dA^a\wedge *dA^b +{\mathcal G}\, dV \wedge*dV \\
&&\qquad\quad +{\mathcal M}^{-1}\, DB\wedge*DB +\frac{M}{k}\epsilon_{ab} DB\wedge Db^a\wedge Db^b \nn 
\eea
This action has the Abelian gauge symmetry $U(1)^2\times U(1)_{d-3}$ that acts on the different fields as
\beq
\delta A^a=d\lam^a\, , \quad \delta b^a=k\lam^a\, , \qquad \quad 
\delta V_{d-2}=d\Lambda\, , \quad \delta B=p\Lambda\, ,
\eeq
for arbitrary real functions $\lam^a$ and $(d-3)$-form $\Lambda$.

\section{Dual actions for gauge theories with continuous center} \label{sec:generaldis}

In this section we provide the generalisation of the duality introduced 
in section \ref{dual_Heisenberg} to other gauge groups with continuous center. 
In subsection \ref{sec:genGauge}, we discuss properties of the gauge groups of interest and stepwise 
introduce a $d$-dimensional gauge-invariant action. This requires to include scalar fields 
that transform under such groups. In subsection \ref{gen:duality} we propose the 
duality that replaces the gauge fields parameterizing the center with $(d-3)$-forms. 
As for the Heisenberg group discussed in section \ref{dual_Heisenberg} 
the dual theory has a simpler gauge group but admits additional topological couplings
of Chern-Simons type. 

\subsection{Gauge theories for groups with continuous center} \label{sec:genGauge}

To begin with, we like to introduce the gauge groups to which a duality similar to the 
one of section \ref{dual_Heisenberg} can be applied. 
We denote the considered Lie group by $G$ 
and name its Lie algebra $\mathfrak g$. The Lie algebra generators $\{t_A\}$ satisfy 
\beq
   [t_A,t_B]=f_{AB}^C t_C\ ,
\eeq
with $f_{AB}^C$ being the structure constants of $\mathfrak g$.
We want to focus on a particular class of Lie algebras, with the following property. Split the generators into $\{t_a,t_\a\}$ with $a=1,\dots,r$ and $\a=1,\dots,n$ and assume they satisfy that
\beq\label{const}
f_{\a B}^A=0 \ .
\eeq
This condition states that the subspace spanned by $\{t_\a\}$ is an Abelian ideal $\mathfrak h$ of $\mathfrak g$. We consider 
the maximal subset of generators $\{t_\a\}$ for which \eqref{const} can be satisfied, i.e.~the case where $\mathfrak{h}$ 
is the center of $\mathfrak g$ and $n$ is its dimension. Note that $\mathfrak h$ corresponds to the algebra of $U(1)^n$.
The quotient $\tilde{\mathfrak g}=\mathfrak g/\mathfrak h$ is a Lie algebra spanned by $\{t_a\}$ with structure constants $f_{ab}^c$. One can easily check that if $f_{AB}^C$, constrained by \eqref{const}, satisfies the Jacobi identities, then $f_{ab}^c$ does too. In more fancy terms, $\mathfrak g$ is a central extension of $\tilde{\mathfrak g}$ by $\mathfrak h$. We denote the Lie group associated to $\tilde{\mathfrak g}$ by $\tilde G$. 

An important feature of $\mathfrak g$ is that its Killing form $B_{AB}=f_{AC}^Df_{BD}^C$ is degenerate. 
Using \eqref{const} one finds that 
\beq \label{deg_BAB}
B_{AB}=\left (\begin{array}{cc}
f_{ac}^df_{bd}^c&0\\
0&0\end{array}\right )\ ,
\eeq
where the upper non-vanishing entries 
correspond to the Killing form of $\tilde{\mathfrak g}$. 
%We do not assume anything about it so it could also be degenerate.
The form \eqref{deg_BAB} implies that $B_{AB}$ cannot be simply used as the gauge coupling function for a 
gauge theory with gauge group $G$. Similar to the discussion in section \ref{Heisenberg_gaugetheory} an action 
for a gauge theory with group $G$ has to be more involved. We discuss its construction next.

To build a gauge theory for the group $G$ we first introduce a collection of gauge fields 
$A^A$ with field strength $F^A=dA^A-\frac{1}{2}f_{BC}^AA^B\wedge A^C$.
Setting $F = F^B t_B$ and $A = A^B t_B$, a gauge transformation $U\in G$ can be represented as
\beq\label{trans}
A\ \rightarrow\ UAU^{-1}-UdU^{-1},\qquad \quad F\ \rightarrow \ UFU^{-1}\ .
\eeq
If $U=e^{\lam^At_A}$ we have, for small $\lam^A$,
\ba
\delta A^A=d\lam^A-f_{BC}^AA^B\lam^C,\qquad \quad \delta F^A=-f_{BC}^AF^B\lam^C.
\ea
Note that $F^A$ transforms in the adjoint of the group $G$ and 
we will denote this representation in the previous section 
by $D[U]_A^B$ for $U \in G$. Explicitly, we have 
\beq
   D[U]_A^B = e^{\lambda^C f_{CA}^B} \ ,
\eeq
where $\lambda^C f_{CA}^B$ has to be understood as a matrix 
in the indices $A,B$ that gets exponentiated.

Next we want to introduce the action of the $d$-dimensional gauge theory.
Consider the definition 
\ba
 S^{(d)} =-\int  Q_{AB}F^A\wedge * F^B+\tilde{S}
\ea
where $Q_{AB}$ is a symmetric real matrix and $\tilde{S}$ is gauge invariant by itself. Gauge invariance of the first term implies that $Q_{AB}$ must transform as
\beq \label{Q-trafo}
   Q_{AB}\quad \rightarrow \quad Q_{CD}D^{-1}[U]_A^CD^{-1}[U]_B^D\ ,
\eeq
as already discussed in \eqref{Q_trans}.
%
%\ba\label{gcf}
%\delta Q_{AB}=(Q_{AC}f_{BD}^C+Q_{BC}f_{AD}^C)\lam^D+\mathcal O(\lam^2),
%\ea
%
The Killing form $B_{AB}$ transforms precisely as \eqref{Q-trafo}, but is degenerate for the considered $G$. Therefore, $Q_{AB}$ must depend on extra scalar fields transforming under $G$. %Let us see how this works.\footnote{Here we mention one possible way of fixing this problem but there are others. For instance, it is enough to include scalar fields in $\tilde G$ instead of $G$. In this case the duality works too.} 

Consider adding to the theory a set of scalars $\{\phi^A\}=\{\phi^a,\phi^\a\}$ living in the group $G$. More precisely, we have that $V(\phi)\equiv e^{\phi^At_A}\in G$. Since $G$ is a Lie group, we can always introduce a Riemannian metric on it such that the group of isometries is (at least) $G$ itself. We declare that these transform under the gauge group $G$ as
\ba\label{right}
V(\phi)\quad \rightarrow \quad V(\phi) \,e^{-\Theta\lam}\ ,
\ea
with $\Theta\lam\equiv \Theta_B^A\lam^At_B$ for some constant $\Theta_{B}^A$ that encodes the particular gauging we are introducing. One typically does that by considering the right-invariant forms
\ba
\tilde \eta= dV(\phi)V^{-1}(\phi)
\ea
which are elements in $\mathfrak g$ invariant under \eqref{right} provided $\lam$ is independent of the spacetime coordinates. Since we are interested in local symmetries, we modify the above as
\ba
\eta= dV(\phi)V^{-1}(\phi)+V(\phi)\Theta AV^{-1}(\phi)\equiv \eta^At_A\ .
\ea
where $\Theta A\equiv \Theta_{B}^A A^Bt_A$. This $\eta$ is invariant under \eqref{trans} and \eqref{right} with $\lam=\lam(x)$ if $\Theta$ satisfies that
\be\label{quad}
f_{AB}^D\,\Theta_D^C=f_{EF}^C\,\Theta_A^E\,\Theta_B^F,
\ee
which can be recognized as the quadratic constraint. It ensures that the scalars $\phi^A$ are charged under a subgroup of $G$. In order to dualize the action, we need to make sure that $\eta^a$ does not depend on $A^\a$ which amounts to imposing
\ba\label{assump}
\Theta_{\a}^b=0.
\ea
Then, eq.\eqref{quad} projects to the quadratic constraint on $\tilde{\mathfrak g}$ as well as
\ba
f_{ab}^c\Theta_c^\a+f_{ab}^\b\Theta_{\b}^\a = f_{ef}^\a\Theta_a^e\Theta_b^f.
\ea
This means that if the extension is non-trivial ($f_{ab}^\a\neq0$) and a subgroup of $\tilde G$ is gauged ($\Theta_a^b\neq0$), then we necessarily need to have a gauging along the centre ($\Theta^\a_\b\neq0$ or $\Theta^\a_b\neq0$).

The extra term in the action is therefore
\ba\label{kinscalars}
\tilde S= - \int \mathcal T_{AB}\,\eta^A\wedge *\eta^B,
\ea
which is gauge invariant and contains the kinetic term for the scalars $\phi^A$. Here $\mathcal T_{AB}$ is a scalar product is the Lie algebra $\mathfrak g$ so it is independent of $\phi$. Expanding \eqref{kinscalars} in terms of the fields $\phi^A$ one finds the Riemannian metric mentioned above. However, it is better to leave it the way it is since it makes gauge invariance manifest. Using the properties of $G$ as well as \eqref{assump}, we find that %$\eta^a$ does not depend on $A^\a$ and
\ba \label{etaalpha_gauging}
\eta^\a=d\phi^\a+\frac{1}{2}f_{bc}^\a\, q^{bc}+\Theta^\a_\b A^\b+ (\Theta_a^\a+D[V]_b^\a\Theta_a^b) A^a.
\ea
where $q^{bc}$ is such that $f_{bc}^\a dq^{bc}=f_{bc}^\a\,\tilde \eta^b\wedge \tilde\eta^c$, due to the Maurer-Cartan structure equations.

We are now ready to build the gauge coupling function $Q_{AB}$ using the scalars $V(\phi)$. If we choose
\ba
Q_{AB}=\mathcal S_{CD}D[V]_E^C\,\Theta_A^ED[V]_F^D\,\Theta^F_B
\ea
with $\mathcal S_{AB}$ a real non-degenerate symmetric matrix independent of $V(\phi)$, then the 
kinetic term for the bosons is gauge invariant. Under a gauge transformation \eqref{right} we have that
\ba
D[V]_A^B\rightarrow D[Ve^{-\Theta\lam}]_A^B=D[V]_A^CD[e^{-\Theta\lam}]_C^{B}
\ea
such that $Q_{AB}$ transforms as desired in \eqref{Q-trafo}.\footnote{Using the quadratic constraint \eqref{quad} we find that $D[e^{-\Theta\lam}]_A^B\Theta_C^A=\Theta_A^BD[e^{-\lam}]^A_C$.}

Note that the adjoint representation for the considered groups $G$ satisfies various 
special properties. Using \eqref{const} one readily shows that it takes the form 
\ba\label{adjprop}
D[U]_a^b&=\delta_a^b+f_{cd}^b\lam^c\tilde D[U]_a^d, &  D[U]_\a^b&=0\ ,& \nn \\
D[U]_a^\b&=f_{cd}^\b\lam^c\tilde D[U]_a^d\ , &  D[U]_\a^\b&=\delta_\a^\b\ ,&
\ea
where $D[U]_a^b$ is precisely the adjoint representation of $\tilde G$ and $\tilde D[U]_a^b$ is a matrix that only depends on $f_{ab}^c$ and can be computed explicitly. Notice that $D[U]_A^B$ does not depend on $\lam^\a$ since $t_\a$ generate the centre of $G$.
%This can be seen from the definition of $D[U]_A^B$ in terms of the structure constants $f_{AB}^C$. Also, 
%For small $\lam$ we have $D^{-1}[U]_B^A=\delta_B^A+f^A_{BC}\lam^C+\mathcal O(\lam^2)$ so the equation above reduces to \eqref{gcf}, as desired. 
%
%\ba
%Q_{ab}&=Q^0_{cd}D[V]_a^cD[V]_\phi^d+Q^0_{\a\b}D[V]_a^\a D[V]_b^\b+2Q^0_{\a c}D[V]_a^\a D[V]_\phi^c\\
%Q_{a\b}&=Q^0_{\a\b}D[V]_a^\a  +Q^0_{b\b}D[V]_a^b\\
%Q_{\a\b}&=Q^0_{\a\b}.
%\ea
%
%
Using the properties \eqref{adjprop} we find that also $Q_{AB}$ has a special form, 
\ba
Q_{ab}&=\mathcal S_{cd}D[V]_e^c\,\Theta_a^e D[V]_f^d\,\Theta_b^f+\mathcal S_{\a\b}(D[V]_c^\a\,\Theta^c_a+\Theta_a^\a)(D[V]_d^\b\,\Theta^b_b+\Theta_b^\b)\ ,\\
Q_{a\b}&=\mathcal S_{\a\g}(D[V]_b^\a\,\Theta^b_a+ \Theta_a^\a)\Theta^\g_\b  \,\qquad \qquad Q_{\a\b}=\mathcal S_{\g\rho}\Theta_\a^\g\Theta_\b^\rho\ .
\ea
where we set $\mathcal S_{\a b}=0$ for simplicity.

To summarize, the full action corresponding to the gauge theory for $G$ coupled to the scalars $\phi^A$ is
\ba \label{final_act_gaugetheory}
  S^{(d)} =-\int Q_{AB}F^A\wedge * F^B + \mathcal T_{AB}\,\eta^A\wedge *\eta^B,
\ea
which is invariant under the gauge transformations \eqref{trans} and \eqref{right}
for every $U\in G$.
%
%\ba\label{nonab}
%A&\rightarrow UAU^{-1}-UdU^{-1},\qquad \quad V(\phi)\rightarrow V(\phi) U^{-1}
%\ea
%
 It is useful to split the index range which allows to rewrite the action as
\ba\label{lagr1}
   S^{(d)} &=- \int  \tilde Q_{ab} F^a\wedge * F^b 
                    +\mathcal S_{\a\b} M^\a\wedge* M^\b  
                    + \mathcal T_{AB}\,\eta^A\wedge *\eta^B
\ea
where we have defined
\beq
\tilde Q_{ab}=\mathcal S_{cd}D[V]_e^c\,\Theta_a^e D[V]_f^d\,\Theta_b^f,\qquad \quad  M^\a=\Theta_\b^\a F^\b+(D[V]^\a_c\Theta^c_a+\Theta_a^\a)F^a.
\eeq
The expression \eqref{lagr1} is useful because it splits the kinetic term for the gauge fields in $G$ in two terms that are gauge invariant independently. The first is simply the kinetic term for the gauge fields in $\tilde G=G/U(1)^n$ while the second term is a combination of $A^\a$ and $A^a$ that is gauge invariant by itself since $M^\a$ is invariant. In particular, it depends on the structure constants $f_{ab}^\a$ that encode the information about the extension of $\tilde{\mathfrak g}$ by $\mathfrak h$.

\subsection{Dualization of the action} \label{gen:duality}

Having introduced in detail the $d$-dimensional gauge theories with 
gauge groups $G$, we are now in the position to perform a dualization 
similar to the one of section \ref{sec_dualHeis}. Since we introduced a full 
representation of scalars $\phi^A$ in \eqref{final_act_gaugetheory} we need to 
perform the analogue steps that yielded the actions \eqref{Spar2} and 
\eqref{Sdual2}.

When perfoming the dualization we first note that the precise 
way to do that depends on the rank of $\Theta_\alpha^\beta$.
As seen in \eqref{etaalpha_gauging} this tensor dictates how many 
scalars $\phi^\alpha$ are shift-gauged by the $A^\alpha$ that 
we like to dualize. As we have seen in section \ref{sec_dualHeis} the 
dualization slightly differs for shift-gauging 
and un-gauged vectors. Therefore, in order 
to present the parent action, we will assume that $\Theta_\alpha^\beta$ has maximal 
rank. The dual action, however, is independent of the rank of $\Theta_\alpha^\beta$
and holds generally. We thus dualize $\phi^\alpha$ into $(d-2)$-forms $V_\alpha$ 
and $A^\alpha$ into $(d-3)$-forms $B_\a$. In order to do that 
we propose the following parent action
\ba\label{parent}
S^{(d)}_{\rm par}=- \int  \tilde Q_{ab}F^a\wedge * F^b + \mathcal T_{AB}\,\eta^A\wedge *\eta^B+
                               \cS^{-1\,\a\b} D B_\a\wedge*D B_\b + 2D B_\a\wedge M^\a
\ea
where we defined
\ba
D B_\a=d B_\a+\Theta_\a^\b V_\b\ .
\ea
The independent variables are $\phi^A,A^A,B_\a$, 
and $V_\a$. This parent action is invariant under $G\times U(1)^n$ given by  \eqref{trans} and \eqref{right} together with
\ba
B_\a\rightarrow B_\a-\Theta_\a^\b\lam_\b,\qquad \quad V_\a\rightarrow V_\a+d\lam_\a
\ea
where $\lam_\a$ is an arbitrary $(d-3)$-form. 

One the one hand, computing the equation of motion for $V_\a$ and inserting the result into \eqref{parent} 
we obtain the original action \eqref{lag1}. On the other hand, when 
we use the equations of motion for $A^\a$ we arrive at the dual action
\ba\nonumber
 S^{(d)}_e&= -\int \tilde Q_{ab} F^a\wedge * F^b + \mathcal T_{ab}\,\eta^a\wedge *\eta^b + \mathcal S^{-1\,\a\b}DB_\a\wedge*DB_\b \\
&\qquad \qquad  \qquad \quad + \mathcal T^{-1\,\a\b}\Theta_\b^\rho\Theta_\a^\g dV_\rho\wedge*dV_\g - f_{ab}^\a DB_\a\wedge \eta^a\wedge\eta^b\ . \label{final_dual_action}
\ea
where we have assumed that $\mathcal T_{a\b}=0$ to make the computations simpler. This Lagrangian is invariant under $\tilde G\times U(1)^n_{d-3}$.

This result is a generalization of the actions found in section \ref{sec_dualHeis} for the Heisenberg group $H$.
To make this match precise, note that in the case of the Heisenberg group we have that $f_{ab}^c=0$ 
and $f_{ab}^3=\mu \epsilon_{ab}$ so the adjoint representation reads
\ba
D[V]_a^b&=\delta_a^b, & D[V]_a^3=\mu \epsilon_{ca}\phi^c\ .
\ea
Thus, the gauged right-invariant forms are
\ba
\eta^a&=d\phi^a+\Theta_b^aA^b\,,\\
\eta^3&=d\phi^3+\Theta_3^3A^3 + \frac{1}{2} \mu \epsilon_{ab} \phi^ad\phi^b+(\Theta_a^3+\mu \epsilon_{cb}\phi^c \Theta_a^b) A^a\, .\nn
\ea
Comparing with section \ref{dual_Heisenberg} we find that $\mu = -Mk$ and that the gaugings are 
given by 
\beq
\Theta_a^b=-k\,\delta _a^b\,,\qquad  \Theta^3_a=0\,,\qquad \Theta^3_3=k^2\, .
\eeq
The fields and coupling functions are then identified as 
\bea
\phi^a&=&b^a\,,\qquad \phi^3=-\frac{k^2}{p}b^3\,,\qquad V_3=-pV\,,\qquad B_3=\frac{1}{k^2}B\,,\\
\cS_{ab}&=&k^{-2}\,\cM_{ab}\,,\qquad \cT_{ab}=\cN_{ab}\,,\qquad \cS_{33}=k^{-4}\cM\,.\nn 
\eea

\section{Conclusions} \label{sec:conclusions}

In this paper we studied the gauge theories for gauge groups $G$ admitting a 
continuous $n$-dimensional center $U(1)^n$. We introduced the kinetic terms 
for the gauge fields by appropriately 
coupling the theory to scalar fields charged under $G$. These scalars ensure 
the existence of a positive definite gauge-coupling function and 
allow the settings to have an interesting vacuum structure. 
We have exemplified this fact by gauging the three-dimensional Heisenberg 
group $H$ and studied its breaking to its discrete version $H_\bbZ$ in the vacuum.
Furthermore, we have shown that the gauge fields in the center of $G$ can be dualized 
to $(d-3)$-forms even in the case that $G$ is non-Abelian. In order to perform this duality
one generally has to also dualize some of the scalar degrees of freedom into $(d-2)$-forms 
if they where non-trivially charged under the center gauge fields. 
The resulting dual theory has a smaller vector gauge group $\tilde G = G/U(1)^n$, supplemented by 
a $U(1)^n$ form-field gauge group. Interestingly, for the Heisenberg group $H$ the dual group
$\tilde H = U(1) \times U(1)$ is an Abelian group. However,  the original non-Abelian structure is not 
lost, but reappears in a Chern-Simons like term that depends on the structure constants 
$f_{ab}^\alpha$ for the center and non-center elements. 

Let us note that the structures we find in this work are expected to appear rather universally 
in string theory. To see this, one has to recall that 
the field strengths are typically involving a coupling to lower-degree 
forms. For example, in Type II string theories one has $F_{p+1} = dC_p + C_{p-3} \wedge H_3$, where 
$H_3$ is the NS-NS three-form field strength. By duality the degrees of freedom in 
$C_p$ are mapped to the degrees of freedom in $C_{8-p}$. When working with the 
lower-degree forms the modification of the field strength will be absent, which can hide 
the non-Abelian structure in effective theories upon dimensional reduction. In other 
words the effective theory is the one with $(d-3)$-forms and Chern-Simons terms that 
we found after dualization. Examples of this feature can be found, e.g.~in \cite{Danckaert:2011ju,BerasaluceGonzalez:2012vb,KashaniPoor:2013en,Grimm:2014aha,Grimm:2015ona}. 
The physics then only becomes fully transparent at the level of the Lagrangian
when dualized to the vector formulation. This is equally true when coupling space-time filling 
D-branes to the setting as was recently discussed in \cite{Grimm:2015ona}.

As an interesting generalization one can consider the coupling of 
additional matter charged under the gauge group $G$.
Such matter can be straightforwardly coupled in the vector formulation 
of the theory. The required additional couplings will depend on 
the bare gauge-fields of the center and it seems that a dual formulation with  
$(d-3)$-forms does in general no longer exist. 
In contrast, when recalling the implementation in M-theory compactifications to three dimensions 
and their lift to F-theory \cite{Grimm:2015ona}, one can be tempted to think that  
such a dual description still exists, at least in three spacetime dimensions. 
It would be very interesting to explore this possibility further and 
clarify the application of this duality in F-theory compactifications with 
a charged matter spectrum. This is particular interesting due to the fact that often a non-Abelian discrete 
gauge symmetry remains as a selection symmetry. For example, the use of the 
discete Heisenberg group as a selection symmetry has recently been disscussed in \cite{BerasaluceGonzalez:2012vb,Marchesano:2013ega,Grimm:2015ona}. Such selection rules can 
be of profound physical importance when studying allowed couplings in 
string theory effective actions.

It is also interesting to point out that the supersymmetrization 
of the actions we found in section \ref{gen:duality} can be challenging. 
While the vector formulation seems to admit a rather straightforward 
supersymmetrization, this is not necessarily true for the dual $(d-3)$-form 
formulation. By duality we expect that it exists and it would be 
desirable to give general supersymmetric forms for the action \eqref{final_dual_action} directly. 
Clearly, the actions will depend on the space-time dimension and the number of 
supersymmetries one wants to realize. An interesting example are theories is $d=3$ with 
$\cN=2$ supersymmetry for which preliminary results are recorded in appendix \ref{3dsusy}.

\subsection*{Acknowledgements}
		
We are grateful to Tom Pugh for initial collaboration and insightful discussions.
We also would like to thank Eran Palti for
useful discussions and comments. This work was supported by 
a grant of the Max Planck Society. It was completed at the Aspen Center for Physics, 
which is supported by National Science Foundation grant PHY-1066293.

\appendix

\section{Supersymmetrization for $d=3$ with $\cN=2$} \label{3dsusy}

In this appendix we propose a way to make an Abelian gauging in $d=3$ compatible with $\mathcal N=2$ supersymmetry which is the completion of the Lagrangian \eqref{Sdual2} for $d=3$. In three dimensions, the field $B$ that appears in \eqref{Sdual2} is a scalar and $V$ a vector. Thus, we work with a set of scalars $f^i$ which include both the $b^a$ and $B$ as well as a set of vectors $A^A$ including $A^a$ and $V$. We discuss how to make the non-standard coupling $DB\wedge Db^a\wedge Db^b$ supersymmetric. In order to make supersymmetry manifest we work in $\mathcal N=2$ superspace and follow the conventions introduced in appendix A of \cite{superspace}.

Let us start with a non-linear sigma model given by 
\ba\label{lag1}
S=\int d^2\th d^2\bar\th\, K(Q^i, \bar Q^i)\, ,
\ea
where $K(Q^i, \bar Q^i)$ is the K\"ahler potential for the chiral superfields $Q^i$. In components these are
\ba
Q^i=f^i+\th^\a\psi^i_\a+\th^2F^i+i\th^a\bar\th^\b\p_{\a\b}f^i+\frac{i}{2}\th^2\bar\th^\a\p_{\a\b}\psi^{i}{}^{\b}+\frac{1}{4}\th^2\bar\th^2\square f^i.
\ea
with $\p_{\a\b}=(\g^m)_{\a\b}\p_m$. Let us assume that $K(Q^i, \bar Q^i)$ is such that its Lie derivative along the vectors $X_A=\Theta^i_A(\p_i+\bar\p_i)$ vanishes, namely,
\be
L_{X_A}K=\Theta^i_AK_i+\Theta^i_AK_{\bar i}=0,
\ee
where the quantities $\Theta^i_A$ are real constants. This implies that the K\"ahler metric for $f^i$ has some Abelian isometries given by shifting the coordinates,
\ba \label{const1}
\delta_\lam f^i=\Theta^i_A \lam^A,\qquad \qquad \delta_\lam \bar f^i=\Theta^i_A \lam^A\, ,
\ea
where $\lam^A$ are arbitrary real constants.

In order to gauge such isometries in a supersymmetric way, we need to introduce vector superfields $V^A$ which have an expansion 
\ba
V^A= \th^\a\bar\th^\b A^A_{\a\b}+i\th^\a\bar\th_\a\phi^A+i\th^2\bar\th^\a\bar\lam^A_\a-i\bar\th^2\th^\a\lam^A_\a+\th^2\bar\th^2D^A
\ea
with $A^A_{\a\b}=(\g^m)_{\a\b}A^A_m$. The parameter $\lam^A$ gets replaced by a chiral superfield $\Lam^A$ under which $V^A$ transforms as
\ba\label{local1}
\delta_\Lam V^A=-\frac{i}{2}(\Lam^A-\bar\Lam^A),
\ea
and the rule \eqref{const} becomes
\ba\label{local2}
\delta_\Lam Q^i=\Theta^i_A \Lam^A,\qquad\qquad  \delta_\Lam \bar Q^i=\Theta^i_A \bar\Lam^A
\ea
The first step is to modify the K\"ahler potential to make it gauge invariant. This is achieved replacing \eqref{lag1} by
\ba\label{susykin}
S_{kin}=\int d^2\th d^2\bar\th \,K(M^i, \bar M^i)
\ea
where we defined
\ba
M^i=Q^i-i\Theta^i_AV^A,\qquad \bar M^i=\bar Q^i+i\Theta^i_AV^A.
\ea
These now transform as
\ba\label{eme}
\delta_\Lam M^i=\delta_\Lam\bar M^i=\Theta_A^i\re \Lam^A\, ,
\ea
which ensures gauge invariance of \eqref{susykin}. The bosonic term that arises from the above is (up to total derivatives)
\ba
\mathcal L_{kin}=-K_{i\bar j}D_mf^iD^m\bar f^j-K_{i\bar j}F^i\bar F^j-2i K_ik^i_AD^A+K_{i\bar j}k^i_Ak^j_B\phi^A\phi^B
\ea
where the covariant derivative is $Df^i=df^i-\Theta_A^iA^A$, so only the real part of the scalar field is actually gauged.

As it stands, this is not an interesting theory because the equation of motion for $D^A$ implies that $K_i\Theta^i_A=0$ which, if plugged back into the action, gives the ungauged theory. Thus, we need to introduce extra terms that contain $D^A$ to make it interesting. The usual solution is to introduce supersymmetric Chern-Simons terms
\ba
S_{CS}=\frac{i}{2}\int d^2\th d^2\bar \th\, \Theta_{AB} V^A\bar D^\a D_\a V^B
\ea
with $\Theta_{AB}$ a symmetric constant matrix and $D_\a$ the covariant spinor derivatives (see appendix A of \cite{superspace} for the conventions). The bosonic terms that this produces are
\ba
\mathcal L_{CS}= \frac{1}{2}\Theta_{AB} A^A\wedge F^B-2\Theta_{AB}\phi^A D^B
\ea
which contains a linear piece in $D^A$. One can also introduce kinetic terms for the vectors which produce quadratic terms in $D^A$.

Here we propose a different way to make the theory consistent which does not rely on CS or kinetic terms for the vectors by adding the following D-term couplings
\ba\label{extra}
S_{extra}=\int d^2\th d^2\bar \th\,\left ( T_{ijk}\,{\rm Im} M^i\,\nabla_\a Q^j \bar\nabla^\a\bar Q^k  +\frac{i}{2}R_{Aij}\,{\rm Im} M^i\,{\rm Im} M^j \bar D^\a D_\a V^A  \right ).
\ea
Let us try to understand these couplings. First of all, $T_{ijk}$ is a constant tensor that needs to satisfy $T_{ijk}=-\bar T_{ikj}$ to make $S_{extra}$ real and $R_{Aij}$ is also constant and symmetric in $ij$. We also defined the covariant derivatives 
\ba
\nabla_\a Q^i=D_\a(Q^i-2i \Theta^i_AV^A),\qquad \bar\nabla_\a \bar Q^i=\bar D_\a(\bar Q^i+2i \Theta^i_AV^A)
\ea
which have the nice property that 
\be
\delta_\Lam (\nabla_\a Q^i)=\delta_\Lam (\bar\nabla_\a \bar Q^i)=0,\qquad  (\nabla_\a Q^i)^*=\bar\nabla_\a \bar Q^i.
\ee
Then, due to \eqref{eme} we have that $\delta_\Lam{\rm Im}M^i=0$ which shows that \eqref{extra} is gauge invariant. Although the first term is naively higher derivative, one can see by expanding \eqref{extra} in components that it is not, just like the last term in \eqref{Sdual2}. Finally, in order to being able to remove the auxiliary fields in a consistent way we need to impose that $\Pi^{A}_BR_{Aij}=R_{Bij}$ where $\Pi^{A}_B=\check \Theta^{A}_i \Theta^i_{B}$ with $\check \Theta^{A}_i$ the Moore-Penrose pseudoinverse.

The full expansion of \eqref{extra} in components is rather long so we just point out that it contains the following  couplings
\ba
\begin{split}
T_{ijk}(Df^i+D\bar f^i)\wedge Df^j\wedge D\bar f^k\\
\left (R_{Aij} + {\rm Im}T_{ijk}\Theta_{A}^i \right ){\rm Im}f^i(Df^j+D\bar f^j)\wedge F^A\\
{\rm Re}T_{ijk}\Theta_{A}^i {\rm Im}f^i(Df^j-D\bar f^j)\wedge F^A.
\end{split}
\ea
The first term has exactly the same structure as the coupling in \eqref{Sdual2}, as desired. Now one can add the kinetic terms for the vectors in the standard way to complete the supersymmetrization of \eqref{Sdual2}. However, we note that this is not necessary and the theory makes sense even if we do not include them.

\bibliographystyle{utcaps} 

\providecommand{\href}[2]{#2}\begingroup\raggedright

\end{document}